\documentclass[final]{svjour2}
\usepackage{graphicx}
\usepackage{rotating}
\usepackage{mathptmx}
\usepackage{amssymb,amsmath,bm}
\usepackage[numbers]{natbib}

\makeatletter
\journalname{Journal of Low Temperature Physics}

\bibpunct{}{}{,}{s}{}{,}

\begin{document}
\newcommand{\hdblarrow}{H\makebox[0.9ex][l]{$\downdownarrows$}-}
\title{DC conductivity in an $s$-wave superconducting single vortex system}

\author{E. Arahata$^1$ \and Y. Kato$^2$}

\institute{1:Research Institute of Industry, The University of Tokyo,\\ 4-6-1 Komaba, Tokyo 153-8505, Japan\\
\email{arahata@iis.u-tokyo.ac.jp}
\\2: Department of Basic Science, The University of Tokyo, \\
3-8-1 Komaba, Tokyo 153-8902, Japan\\
\email{yusuke@phys.c.u-tokyo.ac.jp}
}

\date{}
\maketitle


\begin{abstract}
We study dynamics of a two-dimensional $s$-wave superconductor in the presence of a moving single vortex.
Our analysis is based on the quasiclassical theory including the Hall term, generalized by Kita[T. Kita, PRB, {\bf 64}, 054503 (2001)].
We numerically calculate the linear response of a moving single vortex driven by a dc external current in a self-consistent way, in the sense that Dyson equation, gap equation, Maxwell equations and generalized quasiclassical equation are solved simultaneously. 
We obtain Hall conductivity induced by vortex motion using the generalized quasiclassical equation, while we confirm that it vanishes in the conventional quasiclassical equation.  



\end{abstract}

\section{Introduction}

One of the long-standing and unsettled issues in vortex physics is microscopic calculation flux-flux Hall conductivity. Although there exist lots of references on microscopic calculation on Hall effect in vortex states\cite{Kopnin2001, Kopnin2002}, self-consistent calculations that cover both clean and dirty superconductors have not yet been reported. For example, in the pioneering work \cite{Kopnin1976} on microscopic calculation of Hall conductivity of single vortex for clean s-wave superconductor in the Gor'kov formalism, only the contribution of quasiparticles bounded near vortex cores has been taken into account. Self-consistent calculation for vortex system in the Gor'kov formalism is numerically prohibitive because this formalism contains high energy normal-state properties unnecessary to calculate low energy properties. 

What we imply by ^^ self-consistent calculation' is the calculation that yields a set of green function and electromagnetic field satisfying the Dyson equation, gap equation, Maxwell equations as well as equation of motion of Green function.  
Importance of self-consistency in the calculation of vortex dynamics lies, as emphasized by Eschrig et al. \cite{Eschrig1999}, in the fact that the charge conservation is not necessarily guaranteed in non-self-consistent calculations.

%
%
The quasiclassical theory\cite{Kopnin2001,Eilenberge1968,Eliashberg1971,SereneRainer1983} of superconductivity describes low-energy properties and it can be derived from Gor'kov theory by integrating out high-energy and short-distance properties. This theory proved to be very useful for description of many properties in superconducting single vortex systems\cite{Kopnin2001}. Self-consistent linear response of single vortex with respect to ac electric field has been obtained\cite{Eschrig1999} within the Eilenberger-Eliashberg theory\cite{Eilenberge1968,Eliashberg1971} (which we refer to as the ^^ conventional' quasiclassical theory ). This theory is, however, unable to describe the flux flow Hall effect in the mixed state of superconductors. 

To overcome this difficulty, several authors\cite{Kopnin1994,Houghton1998,Kita2001} generalized the quasiclassical theory such that the Hall effects are taken into account. While these generalized quasiclassical theories open a route to self-consistent calculation of Hall effect in vortex states, no reports have been done on self-consistent calculation on the basis of those theories.
   
%

In this paper, we present the results of self-consistent calculation of generalized quasiclassical equation derived by Kita\cite{Kita2001,Levanda1994} and discuss the linear response Hall conductivity of a moving single vortex driven by external current. We also compare our results with those obtained by the conventional quasiclassical equation. 
%
%
%
\section{Model}

We consider two-dimensional s-wave superconductors.
 The quasiclassical theory is formulated in terms of the quasiclassical Nambu-Keldysh propagator $\check g_{\epsilon}({\bf p}_{\rm f},{\bf r},t)$ in Nambu-Keldysh space, and a function of energy $\epsilon$, and momenta ${\bf p}_{\rm f}$ on the Fermi surface, position {\bf r} and time $t$. The Fermi wave vector and velocity are given by ${\bf k}_{\rm f}={\bf p}_{\rm f}/\hbar$, and ${\bf v}_{\rm f}={\bf p}_{\rm f}/m$. Hereafter, we sometimes drop the subscript f in ${\bf p}_{\rm f}$ for convenience.
We denote usual Nambu-Keldysh matrices : 
$
\check g =\left(\begin{array}{cc}\hat g^{\rm R} & \hat g^{\rm K} \\0 & \hat g^{\rm A}\end{array}\right)
$
with 
$
\hat g^{\rm R,A,K} =\left(
\begin{array}{cc}
g^{\rm R,A,K} & f^{\rm R,A,K}\\
-f^{\dagger, {\rm R,A,K}} & \bar g^{\rm R,A,K}\end{array}
\right) 
$\cite{Kopnin2001,Kita2001}.
The transport equation is given by
\begin{eqnarray}
[\epsilon \check \tau_3+\check \sigma, \check g]_\circ+i\hbar{\bf v}_{\rm f}\cdot {\partial_{\bf r}}
\check g 
+ \frac{\hbar}{2} O_g \{\check \tau_3, \check g \} =0.
\label{eq: transportequation}
\end{eqnarray}
Here $\check \sigma \equiv \check \sigma^{\rm imp}- \check \Delta$ is the difference of impurity self-energy in the Born approximation $\check \sigma^{\rm imp}_\epsilon({\bf r},t)$ and the matrix for pair-potential $\check \Delta$. 
The former is given by 
      $
        \check \sigma^{\rm imp}_\epsilon({\bf r},t)= \frac{i\hbar }{2\tau_{\rm n}}\langle\check g_\epsilon ({\bf p},{\bf r},t)\rangle_{\bf p}, \
      $ 
with the relaxation time $\tau_{\rm n}$ in the normal state. We introduce the notation $\langle\cdots\rangle_{\bf p}=\int\frac{d\theta_{\rm p}}{2\pi}(\cdots)$ with ${\bf p}_{\rm f}=(\cos\theta_{\rm p},\sin\theta_{\rm p})$ for the average on the two-dimensional Fermi surface. The notation $\check \Delta$  denotes the matrix 
$
        \check \Delta =
\left(\begin{array}{cc}
\hat  \Delta & 0 \\
 0 & \hat  \Delta\end{array}
\right)\ 
$ whose element is given by
$
\hat \Delta =\left(\begin{array}{cc}
0 & -\Delta \\ 
\Delta^* & 0\end{array}\right)
$
with the pair-potential $\Delta =\Delta ({\bf r},t)$ satisfying the weak-coupling gap equation 
   $ 
     \Delta ({\bf r},t)=N_{\rm f}V \int  \langle f_{\epsilon}^{\rm K}({\bf p},{\bf r},t)\rangle_{\bf p}
\frac{d \epsilon}{4 }. 
    $
      
In eq.~(\ref{eq: transportequation}), $\check \tau_3 = \left(\begin{array}{cc}\hat \tau_3 & 0 \\0 & \hat \tau_3\end{array}\right)$ with $\hat \tau_3 = \left(\begin{array}{cc}1 & 0 \\0 & -1\end{array}\right)$, $[A,B]_{\circ}=A  \circ B-B \circ  A $, $\{A,B\}=AB+BA$. The notation $A  \circ B$ is defined by\\
$
A  \circ B\equiv \exp \left[ \frac{i\hbar}{2} (\partial_\epsilon \partial_{t'} -   \partial_{t} \partial_{\epsilon'}+\partial_{\bf r} \cdot \partial_{\bf p'} -   \partial_{\bf p} \cdot \partial_{\bf r'})\right]A_{\epsilon}({\bf p}, {\bf r},t)B_{\epsilon'}({\bf p}', {\bf r}',t')\Big|_{\epsilon',{\bf p}', {\bf r}',t'\rightarrow \epsilon, {\bf p}, {\bf r},t}.
$
Here the gauge invariant derivatives are given by

\begin{eqnarray}
\partial_t&=&
\frac{\partial }{\partial t},~~~~~~~~~~~~~  \partial_{\bf r}=\frac{\partial }{\partial {\bf r}}~~~~~~~~~~~~~~~~{\rm on} \  g,\ \bar g, \ \ {\bf E}, \ {\bf B}  \\
 \partial_t&=& \frac{\partial }{\partial t} +\frac{2 i e \Phi}{\hbar}, \ \partial_{\bf r}= \frac{\partial }{\partial {\bf r}} - \frac{i 2e{\bf A}}{\hbar} ~~~ {\rm on} \  f, \ \Delta  \\ 
\partial_t&=& \frac{\partial }{\partial t} -\frac{2 i e \Phi}{\hbar}, \ \partial_{\bf r}= \frac{\partial }{\partial {\bf r}} + \frac{i 2e{\bf A}}{\hbar} ~~~
{\rm on} \  f^\dagger, \ \Delta^*. 
\end{eqnarray}
Here the charge unit $e$ is taken to be negative. We denote the vector potential ${\bf A}$ and the scalar potential $\Phi$, which are related with electromagnetic fields as                 
${\bf E}=-\nabla \Phi-\frac{\partial A}{\partial t},$ 
${\bf B}=\nabla \times {\bf A}$.   
%
%
%
In the last term in the right-hand side of eq.~(\ref{eq: transportequation}), which we call ^^ the Hall term' in the following part, we introduce the notation \cite{Kita2001}
\begin{eqnarray}
O_g&=&e({\bf v}_{\rm f}\times {\bf B})\cdot \frac{\partial }{\partial {\bf p}}+e{\bf v}_{\rm f}\cdot {\bf E}\frac{\partial }{\partial \epsilon}.   
\label{O_g}
\end{eqnarray}
      Normalization condition on the green function is expressed in terms of the local density of state $N({\bf r},\epsilon)$ as 
$N({\bf r},\epsilon)/N_{\rm f}\rightarrow 1$ at $|\epsilon/\Delta_{\infty}| \gg1$, where 
\begin{equation}
N({\bf r},\epsilon)=\frac{N_{\rm f}}{2} \langle[ g_{\epsilon}^{\rm R}({\bf p},{\bf r})- {g}_{\epsilon}^{\rm A}({\bf p},{\bf r})]\rangle_{\bf p}, 
\label{eq: LDOS}
\end{equation}
with the density
of states $N_{\rm f}$ at the Fermi surface in the normal state; $N_{\rm f}=|{\bf p}_{\rm f}|/(2\pi \hbar^2 |{\bf v}_{\rm f}|)$ for two-dimensional system of particles with the paraboric dispersion. $\Delta_\infty$ denotes  the modulus of the pair-potential in the bulk.
We also solve the Maxwell equation 
$\nabla \cdot {\bf E}=\frac{\rho}{\varepsilon}$,  
$\nabla \times {\bf B}=\mu {\bf j}$.
The current and charge density around the vortex are given by
\begin{eqnarray}    
{\bf j}({\bf r},t) =eN_{\rm f} \int \langle{\bf v}_{\rm f} [ g_{\epsilon}^{\rm K}({\bf p},{\bf r},t)- {\bar g}_{\epsilon}^{\rm K}({\bf p},{\bf r},t)]\rangle_{\bf p}\frac{d \epsilon}{4 }
\\
 \rho({\bf r},t) =e N_{\rm f}\int\langle[ g_{\epsilon}^{\rm K}({\bf p},{\bf r},t)+ {\bar g}_{\epsilon}^{\rm K}({\bf p},{\bf r},t)]\rangle_{\bf p}
\frac{d \epsilon}{4 }.
\end{eqnarray}
The scalar electric potential does not appear in the electron density in terms of the gauge-invariant Green function \cite{Kopnin_Book}. However the electron density involves the scalar electric potential and the effect of screening which leads to the condition of local charge neutrality.  
Thus, we approximate the permeability $\mu$ and dielectric constant $\varepsilon$ as those in vacuum $\mu_0$ and $\varepsilon_0$.


In the linear response, we split 
 the physical quantities ${\cal O}(=\check g, \check \Delta, \check\sigma^{\rm imp}, {\bf A}, \Phi)$ into an unperturbed part and a term of first order in the perturbation\cite{Kopnin1994,Kopnin2001}
            \begin{eqnarray}
        {\cal O}({\bf r},t)={\cal O}_0({\bf r},t)+\delta{\cal O}({\bf r}).
          \end{eqnarray}

We solve the generalized quasiclassical equation in a self-consistent way numerically. The numerical procedure of calculating the generalized quasiclassical equation is similar to a self-consistent calculation described in Ref. 4. The self-consistent solution of the quasiclassical equations ensures the conservation law of the charge density $ \partial_t \rho+\nabla \cdot {\bf j}({\bf r})=0$ within the linear response.
We take the quasiclassical parameter $k_{\rm f}\xi=50$ with the coherent length $\xi=\hbar v_{\rm f}/(\pi \Delta_\infty)$. We also take the impurity scattering rate in the normal state $\Gamma_{\rm n}=0.1\Delta_\infty$,with $\Gamma_{\rm n}=\frac{\hbar}{2\pi\tau_{\rm n}}$.
In this paper, we choose a low temperature $T=0.3T_{\rm c}$ and ${\bf v}_{\rm v}=|{\bf v}_{\rm v}| \hat x$ to see the Hall effect clearly.
\section{Results}
\subsection{Equilibrium case}
\begin{figure}[htbp]
\begin{center}
\includegraphics [width=\textwidth] {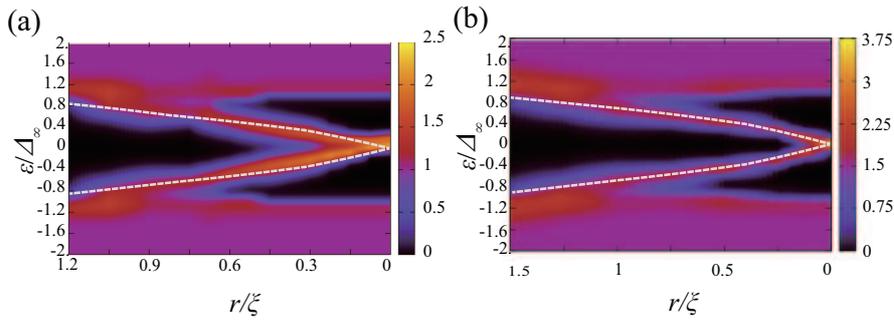}
\end{center}
\caption{(Color online) Local density of states for the generalized quasiclassical equation (a) and the conventional quasiclassical equation (b). The dotted lines represent the dispersion relation eqs.~(\ref{eq: CGM-dispersion}) and (\ref{eq: CGM-dispersion-b}).
\label{fig: ldos-equilibrium}
}
\end{figure}

Figure \ref{fig: ldos-equilibrium}(a) shows LDOS (\ref{eq: LDOS}) in the equilibrium case for the generalized quasiclassical equation including the Hall term. For comparison, we show LDOS for the conventional quasiclassical equation in Fig.~\ref{fig: ldos-equilibrium}(b). 
Using the generalized quasiclassical equation, we obtain asymmetric LDOS as a function of $\epsilon$ in contrast to the conventional equation.

Low energy excitations near a single vortex are exhausted by the Caroli-de Gennes-Matricon mode\cite{Caroli}, or the Andreev bound states\cite{Kopnin2001,KramerPesch74,PeschKramer74}, the dispersion of which is given as a function of impact parameter $r$,

\begin{equation}
\epsilon=E(r)\equiv \frac{r}{C}\int_0^\infty \frac{|\Delta_0(s)|}{\sqrt{r^2+s^2}}{\rm e}^{-u(s)} d s,\quad C\equiv \int_0^\infty {\rm e}^{-u(s)} d s,
\label{eq: CGM-dispersion}
\end{equation}
with 
\begin{equation}
u(s)\equiv\frac{2}{\hbar |{\bf v}_{\rm f}|}\int_0^{|s|}|\Delta_0(s')|d s'.
\label{eq: CGM-dispersion-b}
\end{equation}

In Fig.~\ref{fig: ldos-equilibrium}(a)(b), the overall peak structures in the LDOS can be fitted by the dispersion Eq.~(\ref{eq: CGM-dispersion}). 
This observation confirms validity of our calculation. 

By a close inspection of Fig.~\ref{fig: ldos-equilibrium} (a), we can see that the main peak at the core in the generalized quasiclassical equations is situated on the positive energy side, in contrast to (b). 
This shift of the main peak originates from the Hall term. Similar shifts have been obtained by STS measurements \cite{Kaneko2012,Hanaguri2012}.
\begin{figure}[htbp]
\begin{center}
\includegraphics [width=0.85\textwidth] {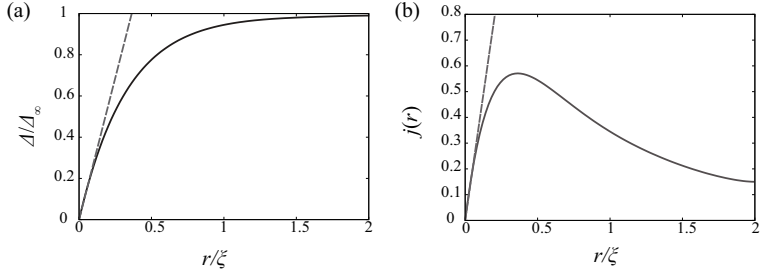}
\end{center}
\caption{
Radial profiles of the modulus of the pair-potential (a) and circular current density (b). Lines represent Eq.~\ref{eq: Delta-KP} scaled by $j_0=2|e|N_{\rm f} v_{\rm f} \Delta_\infty$.}
\label{fig: pair-potential-circular-current}
\end{figure}

Another check of validity can be done by examining the spatial dependence of the pair-potential and the circular current density near the vortex center. Figure~\ref{fig: pair-potential-circular-current} shows the modulus of the pair-potential and the circular current density as functions of the radial coordinate from the vortex center. We see that the spatial dependence near the vortex center is steep. This shrinkage of vortex core at temperatures much lower than the transition temperature was found and discussed by Kramer and Pesch\cite{KramerPesch74}; we can obtain the initial slope of $|\Delta_0(r)|$ and that of the modulus of the current density, respectively, as 
\begin{equation}
|\Delta_0(r)|\rightarrow \frac{\pi \hbar |{\bf v}_{\rm f}| N_{\rm f}V}{8C}\frac{r E'(r)_{r=0}}{2k_{\rm B}T},\quad
|\bm j(\bm r)|\rightarrow \frac{\pi\hbar N_{\rm f} |e| |{\bf v}_{\rm f}|^2}{8C}\frac{r E'(r)_{r=0}}{k_{\rm B}T}
\label{eq: Delta-KP}
%
\end{equation}
along a calculation similar to that in Ref.~13.
Here $V$ denotes the strength of attraction in the weak-coupling regime,
$
(N_{\rm f}V)^{-1}=\ln\left(\frac{T}{T_{\rm c}}\right)+\sum_{m=0}^{\epsilon_{\rm c}/(2\pi k_{\rm B}T)}\frac{1}{m+1/2},
$
with the cut-off frequency $\epsilon_{\rm c}$.
We see that the relations (\ref{eq: Delta-KP}) hold in Fig.~\ref{fig: pair-potential-circular-current}.

\subsection{Linear responses}
\begin{figure}[htbp]
\begin{center}
\includegraphics [width=0.85\textwidth] {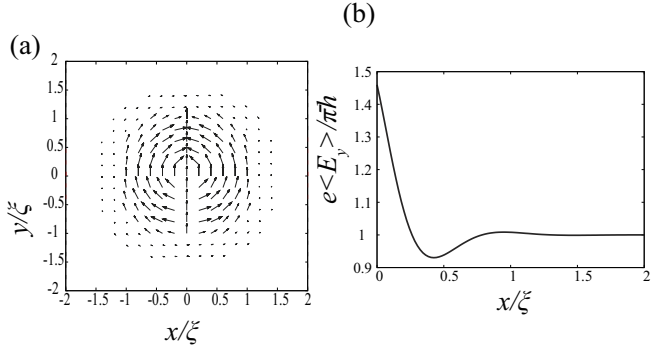}
\end{center}
\caption{
(a) Electric field distributions for the generalized quasiclassical equation. (b)  Radial profiles of the ^^ average' electric field, $\langle E_y\rangle_r$.}
\label{fig: linear-response-electric-field}
\end{figure}

In Fig.~\ref{fig: linear-response-electric-field}, we show electric field distributions for the generalized quasiclassical equation and radial profiles of the ^^ average' electric field, $\langle E_y\rangle_r =|\int E_y dS |/(|{\bf v}_{\rm v}|)$. We can check that the Josephson relation in the flux flow state $\langle {\bf E}\rangle =\langle{\bf B}\rangle \times {\bf v}_{\rm v}$\cite{Josephson} holds, where $\langle \quad\rangle$ denotes the spatial average. The Josephson relation in the present case reduces to $\lim_{r\rightarrow \infty}\langle E_y\rangle_r =\pi \hbar/|e|$. We see that this relation holds in Fig.~\ref{fig: linear-response-electric-field}(b). 

\begin{figure}[htbp]
\begin{center}
\includegraphics [width=0.85\textwidth] {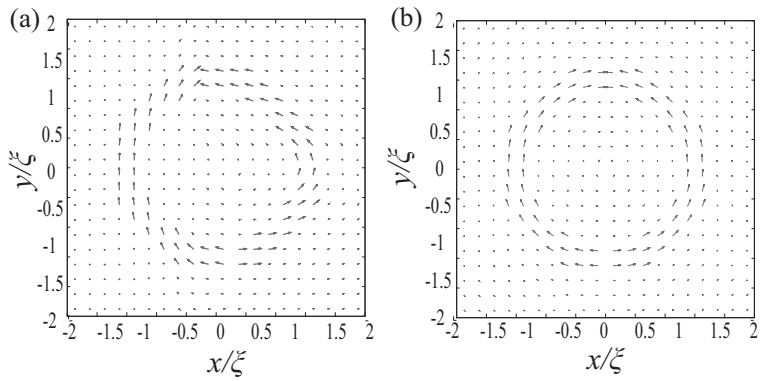}
\end{center}
\caption{
 Current distributions for the generalized quasiclassical equation (a) and the conventional Eilenberger-Eliashberg equation (b). The velocity of vortex is taken to be parallel to the $x$-direction. }
\label{fig: linear-response-current}
\end{figure}

Figure~\ref{fig: linear-response-current} shows the spatial distribution of current density induced by vortex motion for the generalized quasiclassical equation (a) and the conventional Eilenberger-Eliashberg equation (b). We see that there exists the Hall current density in (a) while there does not in (b).  
%
%
We also confirm that the current density ${\bm j}(\bm r)$ approaches a constant vector ${\bm j}_{\rm tr}$ far away from the vortex core (${\bm j}_{\rm tr}$ is identified as the transport current density).
We can thus obtain the flux flow ohmic (longitudinal) $\sigma_{\rm O}=j_{{\rm tr},y}/\langle E_y\rangle$ and Hall conductivities $\sigma_{\rm H}=j_{{\rm tr},x}/\langle E_y\rangle$ using the generalized quasiclassical equation and obtain $\sigma_{\rm O}\simeq 0.68\sigma_B$ and $\sigma_{\rm H}\simeq -0.32 \sigma_B$ with $\sigma_B=N_{\rm f}p_{\rm f}v_{\rm f}|e|/B$. We confirm that the generalized quasiclassical equation captures the Hall effect. 
\section{Conclusion}

On the self-consistent numerical calculation of the generalized quasiclassical equation,
we have observed (i) asymmetric local density of states and the shift of the main peak at the core and (ii) the Hall effect.
We also checked that these results (i) and (ii) originate from the Hall term and are not obtained by the conventional quasiclassical equation.

The ^^ original' generalized quasiclassical equation Eq.~(54) in Ref.~\cite{Kita2001} has an extra term $O_f$ and $O_g$ has a more complicated form. 
In our model, we assume that ${\bf E}$ and ${\bf B}$ vary slowly in time, 
we then have that $O_f\simeq 0$ and Eq.~(\ref{O_g}). 
This assumption is expected to be valid in the calculation of dc conductivity.    
In this paper, we chose the quasiclassical parameter $k_{\rm f}\xi=50$ and fixed $\Gamma_{\rm n}=0.1\Delta_\infty$ to see the Hall effects clearly.
In this case, we found that $O_f$ and the difference between the ^^ original' $O_g$ and Eq.~(\ref{O_g}) are negligibly small (less that 3\%).
For most of actual type-II superconductors, the quasiclassical parameter should be $k_{\rm f}\xi=1000 \sim10000$. 
The calculation for the system with $k_{\rm f}\xi\sim1000$ is thus a future problem.
%

\begin{acknowledgements}
E. A. is supported by a Grant-in-Aid from JSPS (238521) and ``Office for Gender Equality, The University of Tokyo" from JST. 
This work is supported by KAKENHI (70183605) from JSPS.
\end{acknowledgements}

\end{document}